# A Brief Status of the Direct Search for WIMP Dark Matter

David B. Cline

Astroparticle Group, Physics Dept., University of California Los Angeles

**Abstract**

Recently, in February 2014, we held a comprehensive meeting at UCLA on the Search for Dark Matter. 190 scientists came to the meeting, many of the leaders in the field of WIMP Dark Matter searches. We first review the data from LUX that excludes the low-mass WIMP region and slightly lowers the XENON100 limits. We provide a brief review of the problems with the claimed low-mass signals. We discuss the current expectations for SUSY-WIMP Dark Matter and show why very massive detectors like Darwin may be required. We discuss some theoretical predictions from the meeting.

# **Contents**

**Introduction**

The search for the origin of Dark Matter is one of the key problems in science. Recently, in February 2014, a comprehensive symposium was held at UCLA to discuss most of the experiments (current and future proposals) to search for Dark Matter. In this paper we only discuss the direct search for Dark Matter particles. At the UCLA meeting some evidence was presented for Dark Matter processes in our galactic center. To learn more about this one can go to the website of the meeting and find the session devoted to the 'Search for Signals in Fermi-LAT for Dark Matter'.

**1. Summary of world limit on low mass WIMP signals: The Search for low mass WIMPs**

With the discovery of a 125 GeV particle by CMS and Atlas that is widely believed to be the Higgs boson, various models of supersymmetric WIMPs increase the expected mass to the 500 GeV or greater and cross-sections to between 10-45 to 10-47 cm2 .

The likelihood of a supersymmetric low mass WIMP from the theory is very remote. Nevertheless claims from DAMA, CoGeNT, and CRESST have not been withdrawn. This is an unfortunate problem in the worldwide search for dark matter particles. At the recent Dark Matter symposium at Marina del Rey there was very strong evidence put forth to limit the possibility of low mass WIMPs. In particular the null CDMS II search for annual variation in the low mass range coupled with the latest XENON100, 225 day exposure strongly constrains the low mass WIMP hypothesis. Recently CDMS has claimed a low-mass signal as well (see limits from CDMS in Fig. 3).

In Figure 1 we show a summary of the current limits on the low mass WIMP region. We note that the CDMS II, Simple, XENON10 limits come from very different methods. Because the claimed cross-section is so large, these methods are all very robust [3][4].

The limits from XENON100 deserve a special discussion. Both the 100 day XENON100 exposure and the more recent 225 day exposure are inconsistent with a low mass WIMP to the 90 percent confidence level. These data are totally independent and not summed in Figure 1. One could assume that the new 225-day data logically reinforce the 100-day limit. There are then five limits: Simple, XENON10 (S2), CDMS II, XENON100, 100 days, and XENON 225-day limits. All are independent and are 90 percent confidence level null limits. We note that the DAMA results are reported as 3σ limits (see Reference 4 for references).

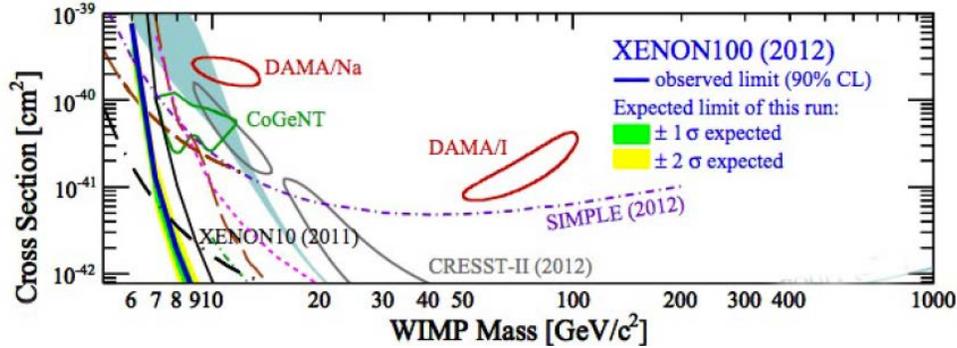

Figure 1. An enlargement of the low mass scale of WIMP searches from the recent XENON100 225 day paper (E. Aprile et al, "Dark Matter Results from 225 Live Days of XENON100 Data," http://arxiv.org/abs/1207.5988).

## 1a. Recent studies of the effect of $^{40}$K in the DAMA experiment

In reference 11 it is shown that the bulk of the singles signal in DAMA is due to radioactive background. Now a new study (see Fig. 3) shows that less than 0.14 Cpd (Fig. 3) can at most be due to WIMPs. This means that the annual variation of the possible WIMPs signal would have to exceed 20%, which is outside any DM model. Reference 12 gives the results presented in Fig. 3. The excellent agreement with Ref. 11 and the excellent fit strongly suggest this is little or no WIMP signal in the data.

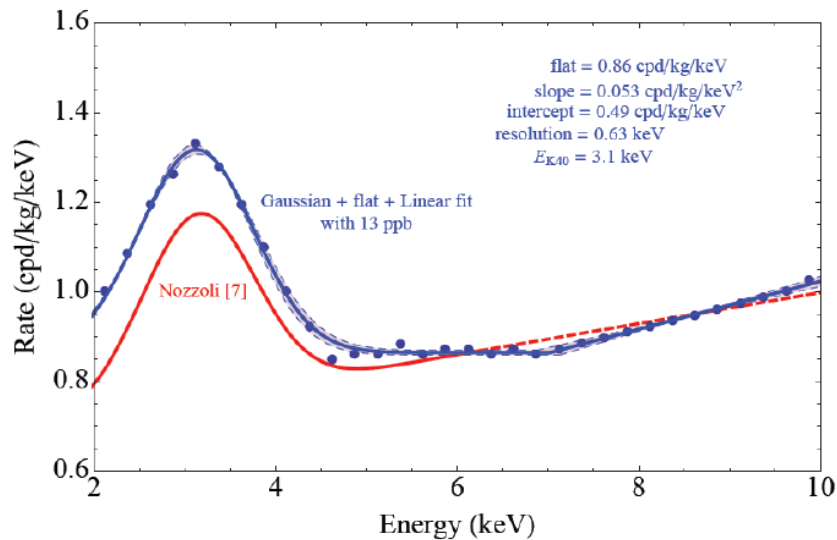

Figure 2. This figure shows the Dama/Libra (dots) and a fit to the data with the correct K (40) and the background from Ref. 11. There is also an estimate by DAMA (red). Even under these circumstances the amount of possible WIMP signal is very low. (Josef Pradler et al, "A reply to criticism of our work (arXiv:1210.5501) by the DAMA collaboration," http://arxiv.org/abs/1210.7548)

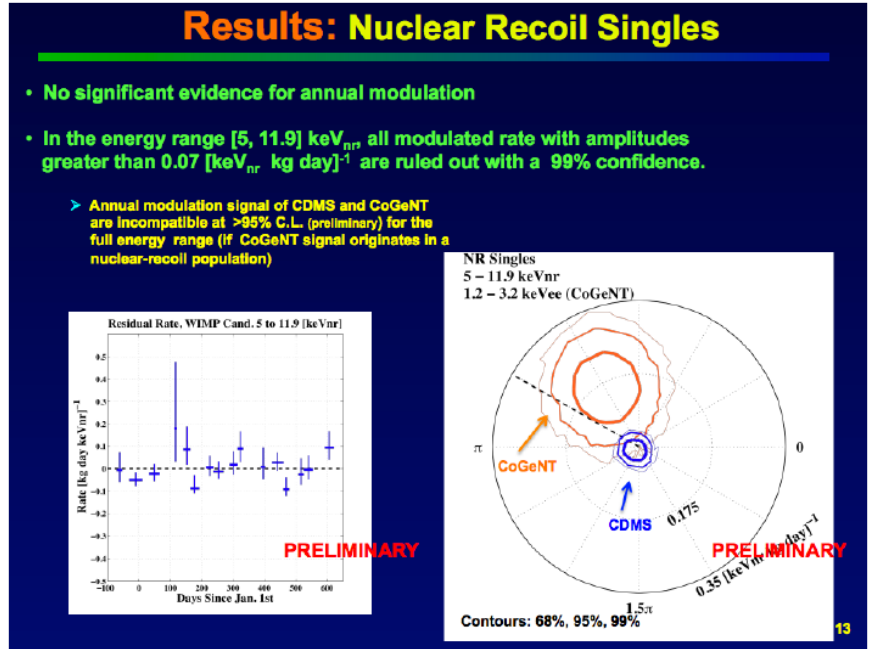

Figure 3. The CDMS group has searched for a low energy signal using the low noise components of the detector as shown in Figure 5. These limits are also shown in Figure 5 (from the CDMS II talk).

These conclusions rule out the DAMA results as a signal for Dark Matter.

## 1b. Neutron signals underground

It is well known that the neutron flux in underground labs has an annual variation. This is likely due to the amount of water or snow in the over burden. In the winter the water absorbs neutrons, in the summer much less so. The ICARUS group measured the LGNS neutron flux as shown in Figure 4. Note that this annual variation fits the DAMA data. DAMA is also at the LGNS. J. Ralston took the ICARUS results and extrapolated over the entire DAMA region (Figure 4) (this is not a fit). Note the excellent agreement with the data. We are not claiming that neutrons make the signal in DAMA, only that there are underground sources that seem to fit the same annual variations than one not due to WIMPs.

## 1c. Signals of Annual Variation underground and DAMA

There are several processes that cause annual variation of processes underground that are similar to the DAMA results.
*1. Radon abundance*
-Has a clear annual increase in the summer and decrease in the winter seen in all underground laboratories
*2. Variation of neutron flux*
-In Figure 5 we show neutron intensity data from ICARUS expanded and compared with the DAMA results. All underground laboratories see a neutron flux annual variation.
*3. The annual variation of cosmic muons as compared with DAMA data (Figure 3)*
-In Figure 4 we show the LVD muon data and compare with the DAMA results. We do not claim a good fit but there is a general agreement.

For all we know DAMA may be seeing a combination of such effects and the phase they observe would be a mixture of these events. Until we identify the actual source of the signals we will not know the actual phase to predict.

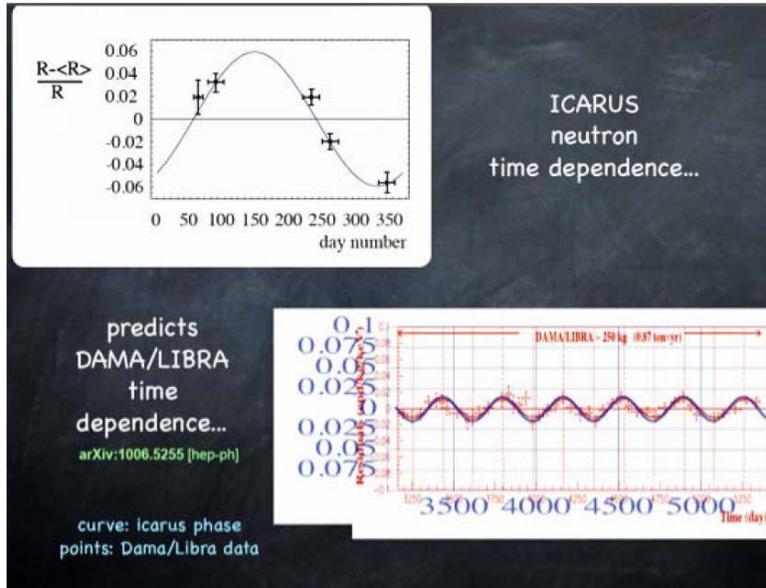

Figure 4. A study of neutron events at the LNGS by the ICARUS group extrapolated to the DAMA results by Ralston (arXiv 1006.5255).

## 2. The LUX results that constrain low mass and high mass regions

At the 2014 UCLA Dark Matter meeting we heard more about the recent results of the LUX detector (all of the following figures and reports can be found in the talks at the February 2014 meeting website - https://hepconf.physics.ucla.edu/dm14/agenda.html). In Figure 5 we show the limits on the LUX results on the whole WIMP region, strongly supporting this paper's introduction. This talk was given by Rick Gaitskell on the LUX results [5].

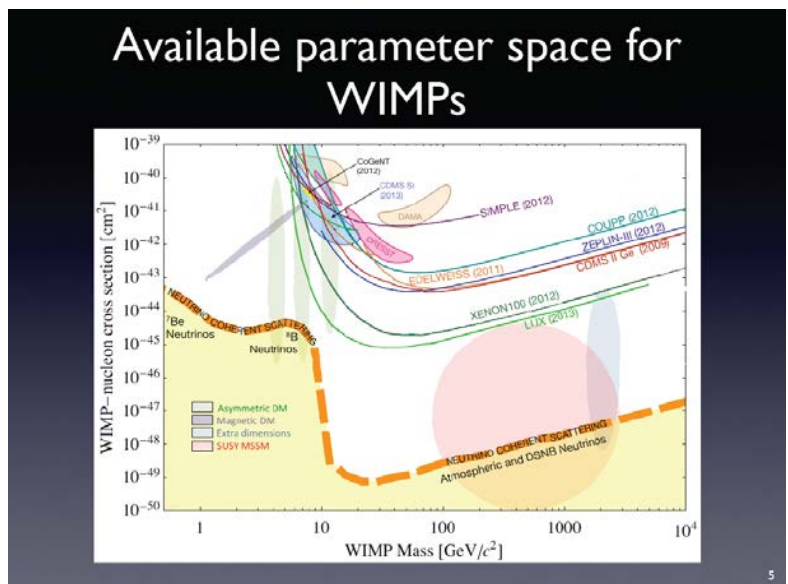

Figure 5. Current limits on WIMP searches [5]. LUX and XENON100 have the lowest limits.

In Figure 6 we show a blow-up of the LUX limits on the low mass WIMP region, strongly supporting the first part of this paper.

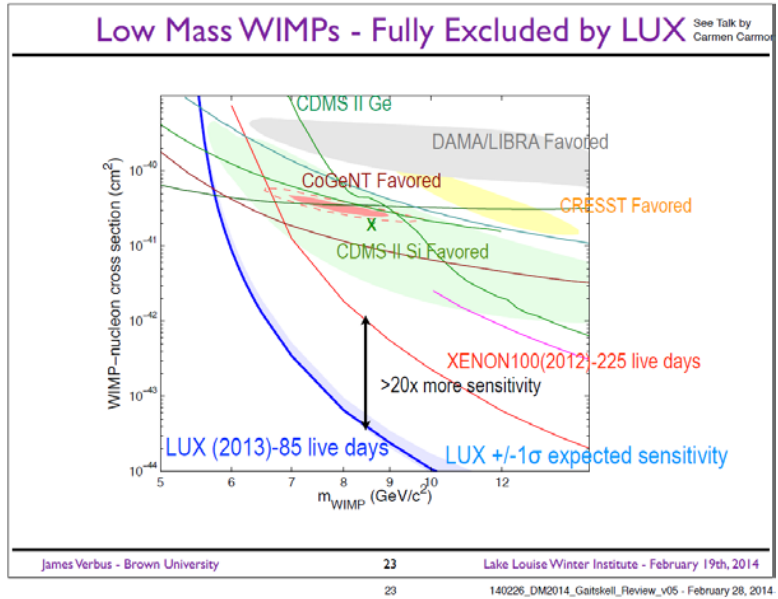

Figure 6. Limit on low mass search from LUX[6] and XENON100.

Figure 7 shows the very large number of events expected in LUX if any of the low mass signals were confirmed. They are not [6]. These results further rule out the DAMA data as being indicative of Dark Matter.

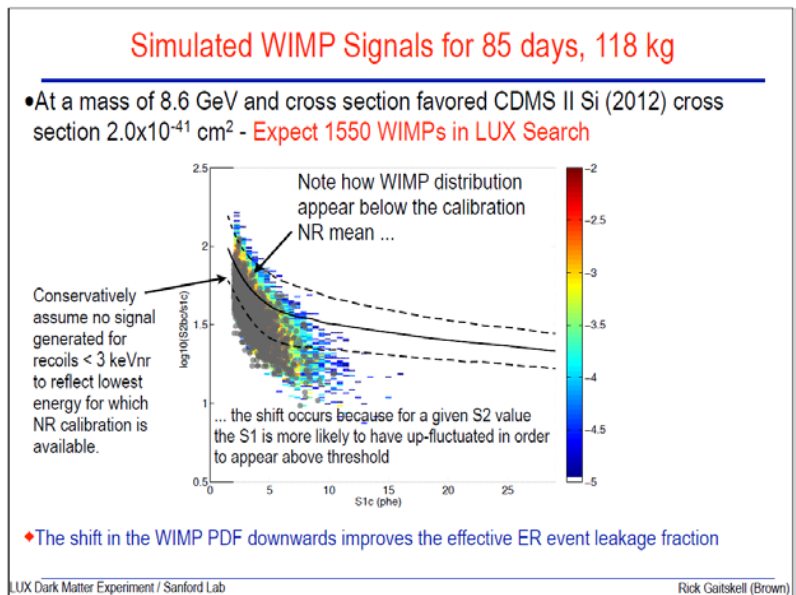

Figure 7. Number of events on LUX if any of the low mass signals are due to Dark Matter [6].

## 3. Some theoretical considerations on the mass of the SUSY WIMP after the discovery of the Higgs boson

Here we rely mainly on the talk of Leszek Roszkowski on the current SUSY theory for the conference[7]. Fig. 8 shows a theory view of Dark Matter concluding that the low mass WIMP region is now excluded and only high mass WIMPs fit the current SUSY model. Fig. 9 shows the impact of the 126 GeV Higgs boson on the Dark Matter sector[7].

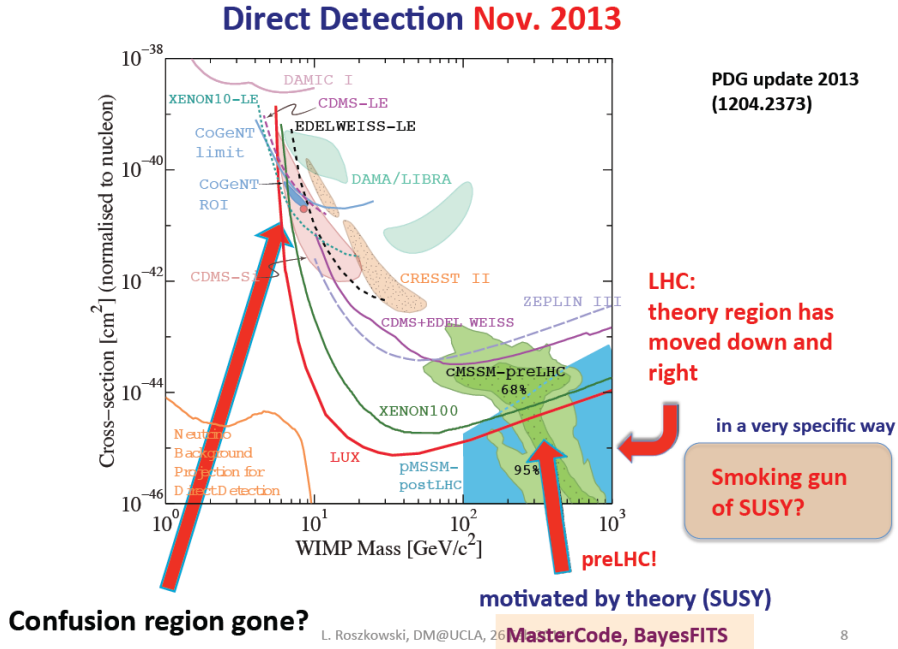

Figure 8. Theoretical regions for viable Dark Matter theory [7].

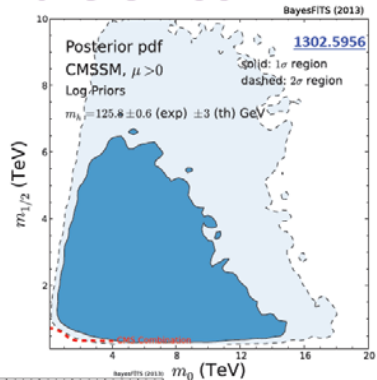
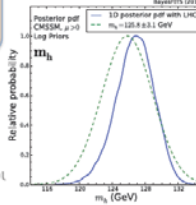

Figure 9. Impact of the Higgs boson on the Dark Matter search [7].

The conclusion of much of the theory talks is that SUSY must be at higher mass, possibly 5 TeV, and that the neutralino could be as massive as 1 TeV. Actually, SUSY may be discovered by the Dark Matter WIMP signal if the squarks are at 5 TeV or more.

**4. Projection of near term and long term WIMP detectors including DARWIN**

With the new limits on squarks from the LHC (~1.5 TeV) and the theory discussed [7] it is clear much larger WIMP detectors are needed [Fig. 5]. Here we limit the discussion to Liquid Noble Gas detectors. Currently we have XENON100 and LUX taking data. The next generation of XENON (1-3) Ton, LZ (7 Ton), Panda X and XMASS (all discussed at DM 2014) will provide the next step. There is now talk of the Generation 3, or G3 detectors. We briefly discuss the DARWIN G3 detector from the talk given by Guiliana Fiorillo at the meeting.

Figure 10 shows the scheme of the 25 Ton DARWIN G3 detector [7].

Figure 10. The DARWIN project (G3 detector proposal). See [8].

If WIMPs are discovered it will be possible to measure the mass and spin using G2 and G3 detectors. An example for DARWIN searching for Dark Matter is shown in Fig. 11.

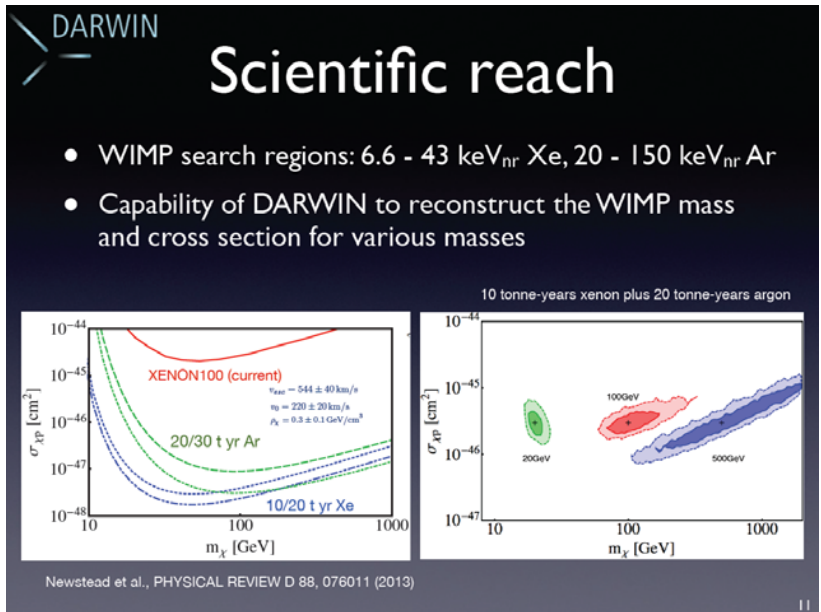
Figure 11. The reach of DARWIN and the possibility of measuring the WIMP mass [8].

At the recent Snowmass meeting the natural neutrino background for WIMPs was determined. This is shown in Fig. 11. It may be possible to go below this limit looking at the momentum distribution of the recoil in a WIMP interaction. Fig. 12 shows the reach of DARWIN [7] that would saturate this background level. The G3 detectors will be needed either to study Dark Matter (Fig. 11) or to make a complete search (Fig. 12). This is why the recent P5 report strongly supported the study of these detectors now.

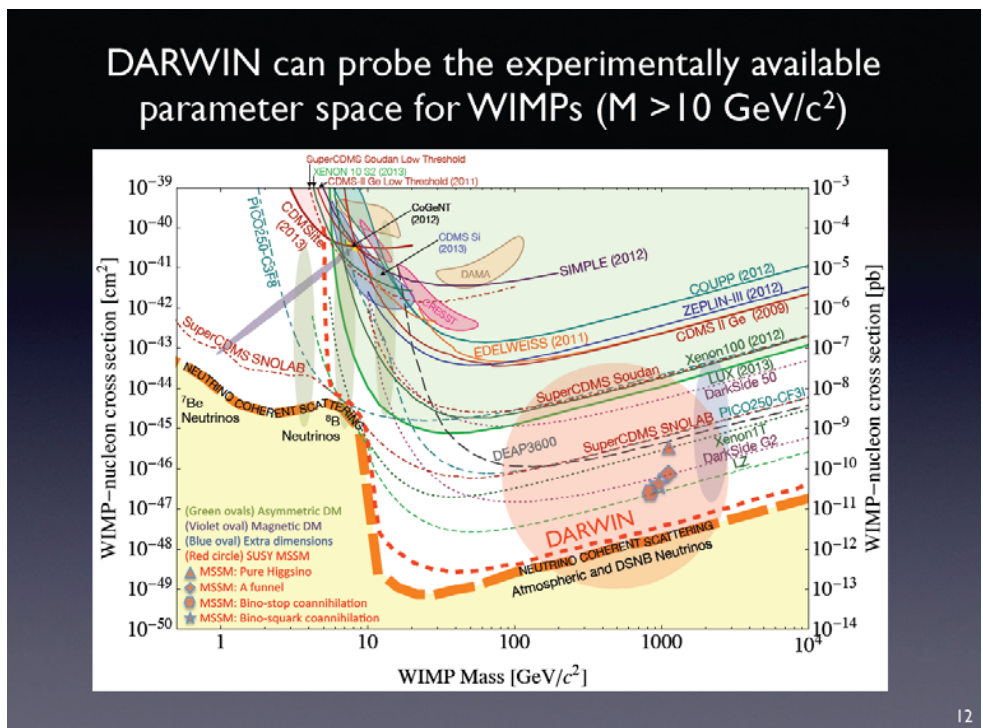
Figure 12. The G3 detectors will be able to search for WIMPs down to the background level [8].

In conclusion, recent developments at the LHC, the Higgs boson discovery, and theory all suggest that the SUSY WIMP may be very massive (up to 1 TeV). For these reasons and the G2 detectors being constructed now, the proposed G3 detectors will be needed to discover Dark Matter.

I would like to thank Elliott Bloom, Elena Aprile, Leszek Roszkowski, Katie Freese, tand the full advisory committee (see website) for help with DM 2014. We will follow this up with DM 2016 in February 2016 at UCLA. I thank Laura Baudis and Giuliana Fiorillo for discussions on DARWIN.

**References**
1. The CMS Collaboration, "Observation of a New Boson at a Mass of 125 GeV with the CMS Experiment at the LHC," arXiv submit/0524428 [hep-ex] 31 Jul 2012.
2. The ATLAS Collaboration, "Observation of a New Particle in the Search for the Standard Model Higgs Boson with the ATLAS Detector," arXiv 1207.7214v2 [hep-ex] 31 Aug 2012.
3. E. Aprile et al, "Dark Matter Results from 225 Live Days of XENON100 Data," http://arxiv.org/abs/1207.5988.
4. D.B. Cline, "The Search for Dark Matter (WIMPs) at Low Mass and with New Methods," Mod. Phys. Lett. A 26 (2011) 925.
5. From the talk of Rick Gaitskell (http://www.pa.ucla.edu/sites/default/files/webform/140226_DM2014_Gaitskell_Review_v06.pdf)
6. From the talk of Rick Gaitskell (http://www.pa.ucla.edu/sites/default/files/webform/140226_DM2014_Gaitskell_Review_v06.pdf)
7. From the talk of Leszek Roszkowski (http://www.pa.ucla.edu/sites/default/files/webform/roszkowski_ucladm_26feb2014_final.pdf)
8. From the talk of Giuliana Fiorillo on the DARWIN Project (http://www.pa.ucla.edu/sites/default/files/webform/fiorillo-DM2014d.pdf)